\documentclass[camera,letterpaper,nomarginnotes,nonarrowgutter]{jpaper}
\usepackage[sort,compress]{cite}
\usepackage{amsmath,amssymb,amsfonts}
\usepackage{graphicx}
\usepackage{textcomp}
\usepackage{xcolor}
\usepackage{xspace}
\usepackage{fancyhdr}
\usepackage{balance}
\usepackage{subcaption}
\usepackage[title]{appendix}
\usepackage{titling}
\usepackage[bookmarks=true,breaklinks=true,letterpaper=true,colorlinks,citecolor=blue,linkcolor=blue,urlcolor=blue]{hyperref}
\usepackage[noabbrev,capitalize]{cleveref}

\newcommand{\TODO}[1]{{\footnotesize\color{red}[TODO: #1]}}

\newcommand{\CRS}[1]{{\color{black}#1}}
\newcommand{\red}[1]{{\color{red}#1}}
\newcommand{\blue}[1]{{\color{black}#1}}
\newcommand{\purple}[1]{{\color{black}#1}}
\newcommand{\Defense}{CnC-PRAC\xspace}
\newcommand{\ignore}[1]{}
\newcommand{\NPRO}{$\text{N}_{\text{PRO}}$\xspace}
\newcommand{\TRH}{$\text{T}_{\text{RH}}$\xspace}
\newcommand{\NBO}{$\text{N}_{\text{BO}}$\xspace}
\newcommand{\ABOACT}{$\text{ABO}_{\text{ACT}}$\xspace}
\newcommand{\ABODELAY}{$\text{ABO}_{\text{Delay}}$\xspace}
\newcommand{\NMIT}{$\text{N}_{\text{mit}}$\xspace}

\newcommand{\ALERT}{\text{Alert}}
\newcommand{\RFM}{\text{RFM}}
\newcommand{\ACT}{\text{ACT}}

\newcommand{\TREFI}{\text{tREFI}}

\newcommand{\REF}{\text{REF}}

\newcommand{\subtitle}[1]{\posttitle{\par\normalfont{#1}\par\end{center}}}

\title{\Defense{}: Coalesce, not Cache, Per Row Activation Counts for an Efficient in-DRAM Rowhammer Mitigation}

\author{Chris S. Lin$^{\dagger}$ \quad Jeonghyun Woo* \quad Prashant J. Nair* \quad Gururaj Saileshwar$^{\dagger}$
\\\emph{$^{\dagger}$University of Toronto} \qquad *\emph{University of British Columbia}\\
shaopenglin@cs.toronto.edu, jhwoo36@ece.ubc.ca, prashantnair@ece.ubc.ca, gururaj@cs.toronto.edu
}

\begin{document}
\bstctlcite{IEEEexample:BSTcontrol}
\maketitle


\thispagestyle{plain}
\pagestyle{plain}



\begin{abstract}
JEDEC has introduced the Per Row Activation Counting (PRAC) framework for DDR5 and future DRAMs to enable precise counting of DRAM row activations using per-row activation counts. While recent PRAC implementations enable holistic mitigation of Rowhammer attacks, they impose slowdowns of up to 10\% due to the increased DRAM timings for performing a read-modify-write of the counter. Alternatively, recent work, Chronus, addresses these slowdowns, but incurs energy overheads due to the additional DRAM activations for counters. 
 
In this paper, we propose \Defense{}, a PRAC implementation that addresses both performance and energy overheads. Unlike prior works focusing on caching activation counts to reduce their overheads, our key idea is to reorder and coalesce accesses to activation counts located in the same physical row. Our design achieves this by decoupling counter access from the critical path of data accesses. This enables optimizations such as buffering counter read-modify-write requests and coalescing requests to the same row. Together, these enable a reduction in row activations for counter accesses by almost \CRS{75\%-83\%} compared to state-of-the-art solutions like Chronus and enable a PRAC implementation with negligible slowdown and a minimal dynamic energy overhead of \CRS{0.84\%-1\%} compared to insecure DDR5 DRAM.

\end{abstract}

\section{Introduction}

Aggressive scaling of DRAM has introduced security vulnerabilities such as Rowhammer (RH)\cite{kim2014flipping}, a data disturbance phenomenon where frequent row activations induce charge leakage and bit flips in adjacent rows. These faults can lead to severe security issues, including privilege escalation and data corruption\cite{seaborn2015exploiting, frigo2020trrespass, gruss2018another, aweke2016anvil, cojocar2019eccploit, gruss2016rhjs, vanderveen2016drammer}. With continued scaling, the Rowhammer threshold (\TRH{}) -- the number of activations needed to induce bit flips -- has dropped from 70K~\cite{kim2014flipping} to 4.8K~\cite{kim2020revisitingRH}, and is expected to decline further, worsening the threat.

To mitigate RH, the DRAM industry has developed a series of in-DRAM defenses, the most recent being Per Row Activation Counting (PRAC)~\cite{jedec_ddr5_prac}, introduced in the DDR5 specification. PRAC addresses both the space and time challenges associated with RH mitigation. It provides dedicated space within DRAM to maintain activation counters for each row, which are incremented on every access. It also enables the DRAM to request mitigation time by asserting an Alert signal to the memory controller, which initiates the Alert Back-Off (ABO) protocol. The controller can issue mitigation commands (RFM) to refresh vulnerable rows through this protocol. PRAC thus establishes a principled interface for RH mitigation. However, current PRAC-based implementations incur performance and energy overheads that hinder widespread adoption.



State-of-the-art PRAC implementations suffer from significant inefficiency due to the read-modify-write operations for updates to counters stored within the DRAM rows. These updates occur every time a row is activated and precharged, causing an increase in the DRAM row precharge latency (tRP) and row cycle time (tRC), and causing a slowdown in overall system performance. Consequently, state-of-the-art PRAC implementations, such as MOAT~\cite{MOAT} and QPRAC~\cite{QPRAC}, incur a slowdown of almost 10\%~\cite{Chronus} compared to an insecure baseline DRAM, due to the increased DRAM timings.

Alternatively, PRAC implementations, such as Chronus~\cite{Chronus}, attempt to mitigate these overheads by reorganizing the activation counter storage. Chronus stores counters in a DRAM sub-array separate from the data sub-arrays, and utilizes sub-array level parallelism~\cite{SALP} to access and update counters in parallel to the data-access, without increasing DRAM timings. While Chronus effectively reduces the performance slowdown associated with PRAC, it still incurs energy overheads. \CRS{Under a closed-row policy,} Chronus requires an additional counter row activation for each data \CRS{row} activation that increases energy consumption. Thus, existing solutions either suffer from high slowdowns or energy overheads that limit their practicality.

In this work, we propose a PRAC implementation to address both performance and energy drawbacks in existing works. 
We build on the Chronus substrate that stores counters in separate sub-arrays from data.
Here, we observe that decoupling counter accesses from the critical path of the data row activations can enable optimizations to reduce counter activations. 
Specifically, we make two observations that frame the design space for counter access optimizations.

\begin{figure*}[!htb]
    \centering
\includegraphics[width=6.7in]{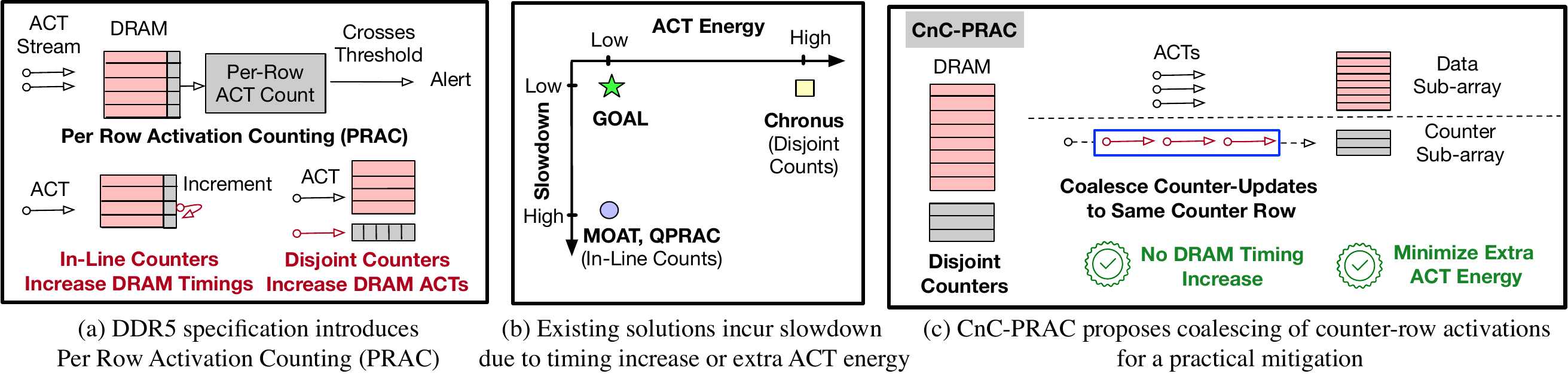}
    \caption{(a) With the PRAC framework, DRAM can store per-row activation counts that can exactly signal when a mitigation is required, when a count crosses a threshold, using an \ALERT{}. (b) Existing PRAC implementations (QPRAC~\cite{QPRAC} or MOAT~\cite{MOAT}) store counts in-line with DRAM rows, and incur increased DRAM timings to increment activation counts; other implementations (Chronus~\cite{Chronus}) store counts in separate sub-arrays, incurring extra activation counts for counters on each data access. (c) We propose \Defense{}, which decouples counter \CRS{row} activations from data \CRS{row} activations, and coalesces counter row activations using a request buffer, avoiding slowdown without DRAM timings increase and minimizing energy for counter row activations.
    }
    \vspace{-0.1in}
    \label{fig:intro}
\end{figure*}

\noindent \textit{\bf Observation-1: Locality in Counter Row Activations.} As each DRAM row (64K rows per bank) requires a 1 byte activation counter (assuming a \TRH of 256), the counters occupy 64KB of storage, fitting in 64 DRAM rows within a chip\footnote{An x8 DRAM chip has 1KB per row, assuming 8KB row per DRAM bank.}. Consequently, there is significant locality among counter row activations, even if data row activations may not have locality. 
This motivates the potential for coalescing requests to the same counter row to reduce counter row activations. 

\noindent \textit{\bf Observation-2: Large Footprint of Counters.} Our analysis shows that across workloads, the number of unique rows that make up even 50\% of the row activations surpasses 7,500 rows per bank. Thus, SRAM counter caches, within DRAM chips, may require 7500+ entries, which can be prohibitive.

Based on these observations, we propose \Defense{}, a RH defense that \underline{C}oalesces, \underline{n}ot \underline{C}aches, \underline{PRAC} counters to minimize counter activations. 
\Defense{}'s main design component is an in-DRAM counter request buffer \CRS{(see \cref{fig:intro}(c))}, which buffers counter requests and is designed to enable coalescing of counter requests to the same row. Each buffer entry tracks the counter sub-array RowID and the byte position of the requested counter. 
When a batch of coalescable entries reaches a threshold (e.g., 4), the request buffer performs a single counter row activation to read-modify-update four counters in the same counter row. These four counter updates to the same row are done in parallel to a data row activation.
 

\CRS{
\Defense{} proposes multiple designs for the request buffer. 
First, we propose one request buffer per counter row, \textit{\Defense{}-PerRow}. This per-row request buffer maximizes coalescing opportunities by independently tracking requests to each counter row, achieving up to 82\% reduction in counter row activations compared to Chronus. This design incurs a higher storage overhead (e.g., 4 entries per row for 64 counter rows incurs 384 bytes per bank), which may be less desirable in space-constrained settings. To reduce this cost, we evaluate an additional design, \textit{\Defense{}-Unified} using a unified request buffer shared across all rows, coupled with a priority-based eviction policy that tracks the row with the highest number of outstanding requests. When four requests accumulate to the same row within this buffer, the buffer flushes them to memory. This reduces counter row activations by 75\%, providing most of the benefits of the per-row design while incurring less storage (e.g., 64 entry unified buffer needs 192 bytes per bank).}

To ensure the counter request buffering does not impact the security against Rowhammer, we ensure that no counter request is buffered for more than $n$ repeated activations at any time. 
We lower the back-off threshold by $n$ (by default, $n=4$) to account for this delay in updating the activation counts in DRAM and thus avoid any loss in security.

We evaluate \Defense{} with 57 workloads, including SPEC CPU-2017, SPEC CPU-2006, TPCC, Hadoop, and YCSB. Our results show that with \CRS{Back-Off threshold (\NBO{})} of 28 and one mitigation per \ALERT{}, \Defense{} can securely handle a \TRH of 66 while incurring a negligible slowdown compared to a non-secure DRAM baseline.
\Defense{} incurs just \CRS{0.84-1\%} extra dynamic energy compared to the baseline.
This is due to a reduction in counter row activations
in \Defense{}. 

\smallskip
\noindent Overall, this paper makes the following contributions:
\begin{enumerate}[leftmargin=*]
    \item We observe that existing PRAC implementations are impractical, either due to high performance or energy overheads imposed by the activation-counter read-modify-writes in PRAC, on the critical path of data-access.
    \item We propose \Defense{}, a decoupled counter management framework in-DRAM to address these overheads.
    \item We propose the first in-DRAM counter coalescing strategies for PRAC that minimize counter row activations.
    \item 
    We achieve a PRAC implementation with negligible performance and \CRS{dynamic} energy costs and no security impact.
\end{enumerate}

\section{Background and Motivation}\label{sec:background}

\subsection{Per Row Activation Counting (PRAC)}
JEDEC's DDR5 specification~\cite{jedec_ddr5_prac} introduces the Per Row Activation Counting (PRAC) framework to count row activations and accurately mitigate Rowhammer attacks. PRAC incorporates two key components: (1) activation counters embedded within each DRAM row, which are incremented upon every activation, and  (2) the \ALERT{} Back-Off (ABO) protocol, where the DRAM signals an \ALERT{} to the memory controller when a row’s counter exceeds a predefined threshold (\NBO{}), to request the memory controller to initiate Rowhammer mitigation commands, Refresh Managements (\RFM{}s), to the DRAM. 

While PRAC provides robust protection against Rowhammer, it requires read-modify-write operations on activation counters during each row activation. This requires an increase in DRAM timings such as row precharge latency (tRP) and row cycle time (tRC), resulting in slowdown and energy overheads.

\subsection{Drawbacks of Prior PRAC Implementations}
Recent approaches such as MOAT~\cite{MOAT} and QPRAC~\cite{QPRAC} focus on providing a secure PRAC implementation, demonstrating vulnerabilities with Panopticon~\cite{bennett2021panopticon}, the precursor to PRAC.
Both MOAT and QPRAC store activation counters inline with the DRAM row, as per the PRAC specification~\cite{jedec_ddr5_prac}, and consequently, pay the cost of the increased DRAM timings for the counter read-modify-write. On average, these approaches incur a performance overhead of up to 10\% compared to insecure DDR5 implementations~\cite{Chronus}.

To address these slowdowns, Chronus~\cite{Chronus} stores the activation counters in a separate DRAM sub-array. This design enables sub-array-level parallelism~\cite{SALP}, allowing counter reads and updates to occur parallel to data \CRS{row} activations. While Chronus effectively reduces performance penalties, it comes at the cost of energy efficiency as each data \CRS{row} activation triggers an additional counter row activation; the prior work reports that it can cause over 10\% energy overhead~\cite{Chronus}. This paper seeks to optimize the management of counters in Chronus to reduce the energy of counter row activations.

\section{Design of \Defense{}} \label{sec:design}
We build on Chronus's substrate, with a separate sub-array for counters compared to the data, to enable parallel activations of counter rows with data row activations. To minimize the extra counter \CRS{row} activations, we decouple the counter access from the critical path of the data access to enable optimizations such as counter row activation coalescing. Next, we first motivate the potential for coalescing before providing details on our counter request buffer design and operation.

\subsection{Potential for Counter Activation Coalescing} 

\begin{figure}[htb]
\centering
\includegraphics[width=3.4in,height=\paperheight,keepaspectratio]{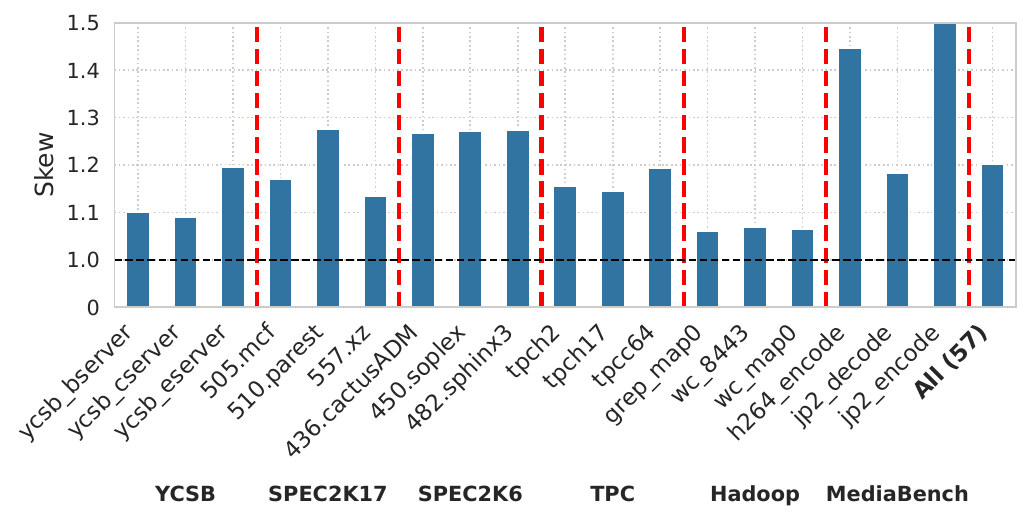}
\caption{The skew (max/mean) of accesses across the 64 rows of the counter subarray. We show the top-3 most-skewed workloads across the benchmarks, observing an average skew of 1.2 across all workloads and a maximum skew of 1.5.}
\label{fig:counter_row_skew}
\end{figure}

\begin{figure}[htb]
\centering
\includegraphics[width=3.4in,height=\paperheight,keepaspectratio]{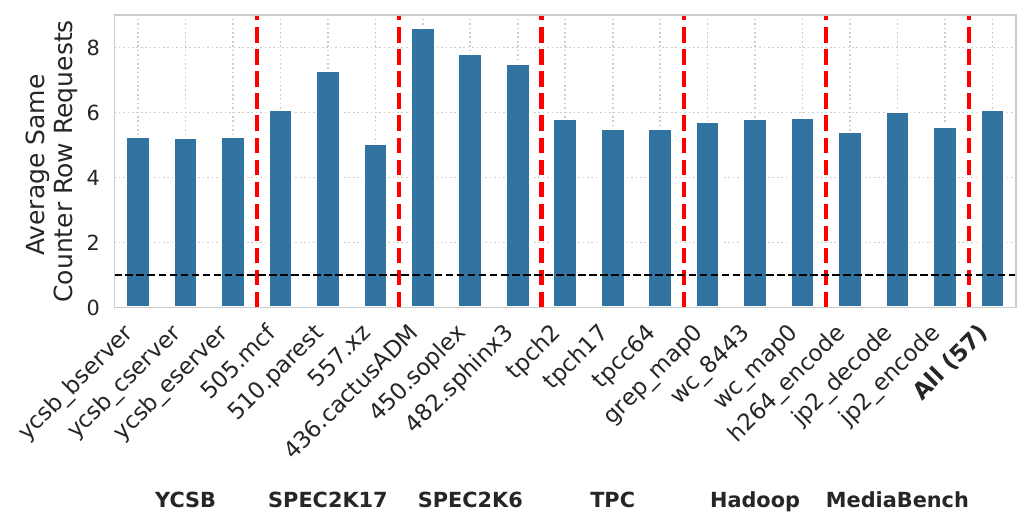}
\caption{The highest number of same row counter requests within any window of 64 counter requests, averaged across windows. On average, across all workloads, within 64 requests, we find 6 requests are to the same counter row.}
\label{fig:same_row_request_avg}
\end{figure}

\CRS{
We analyze both the spatial distribution and temporal locality of counter row activations to evaluate the feasibility of coalescing these accesses.

\cref{fig:counter_row_skew} shows the skew in activations across the 64 counter rows in a DRAM bank, defined as the ratio of maximum to mean activations. On average, there is a skew of 1.2, with the worst-case of 1.5 in Mediabench. A low spatial skew implies that accesses may be distributed evenly, making coalescing via row buffer reuse ineffective. That is, counter row accesses rarely hit the row buffer naturally and typically require an activation. 

To better understand coalescing potential, we also analyze the temporal repetition of counter row accesses. 
In \cref{fig:same_row_request_avg}, we plot the highest number of requests to the same counter row within a window of 64 requests, averaged across all such windows, for a workload. We find that within a window of 64 counter requests, 6 requests are to the same counter row, on average across all workloads. This suggests that, although workloads have low spatial skew on average, short-term temporal locality exists, making it feasible to coalesce multiple accesses to the same row. This motivates our counter request buffer, which buffers up to $n$ counter requests, to coalesce multiple requests into a single counter row activation.

The coalescing potential increases as the request buffer size increases. In \Defense{}, we aim to minimize the request buffer size, to enable coalescing even with a buffer size of 64 (equivalent to the number of counter rows in the sub-array).
}

\subsection{Request Buffer \CRS{Operations}} 

\begin{figure}[htb]
 \centering
\includegraphics[width=\textwidth/2,height=4cm,keepaspectratio]  {"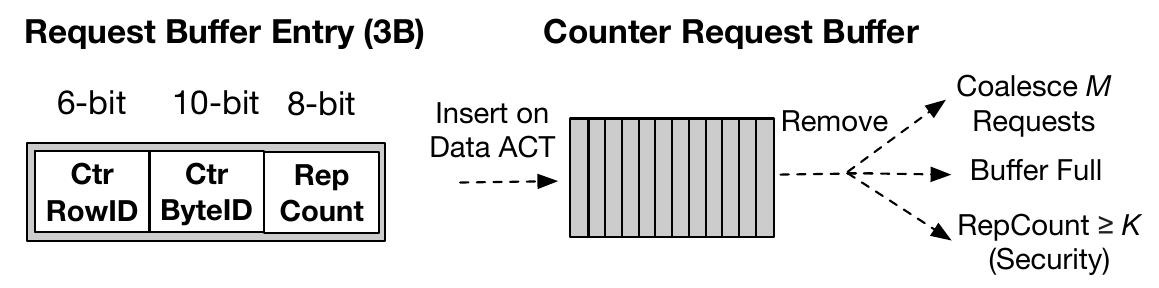"}
  \caption{Design of \Defense{}. Any data \CRS{row} activation inserts a counter request into the buffer. An entry is removed when $M$ requests to the same row can be coalesced, if the buffer is full, or if it requires removal for security.}
  \label{fig:rb-design}
\end{figure}

\noindent \textbf{\CRS{Overview}:} The request buffer is an $n$-entry structure implemented using content-addressable memory (CAM). Each entry is 3 bytes, comprising a 6-bit counter subarray row identifier (RowID), a 10-bit byte-position ID (ByteID) to identify a specific counter, and the remaining 8 bits are used to count repeated requests (RepCount) for the same counter while the request is buffered. The RowID serves as the key, enabling counter row activation coalescing for requests to the same counter row. 
We describe the request buffer operation next.

\smallskip
\noindent \textbf{Insertion:} On each data row activation, instead of directly accessing the counter from the sub-array, we insert the counter request into the buffer, as shown in \cref{fig:rb-design}. For new entries, the RepCount is set to zero. If there is already a request for the same counter, we increment the corresponding RepCount instead of allocating a new entry.

\smallskip
\noindent \textbf{Removal:} Accesses to the counter sub-array for counter read-modify-writes of counters are done on request buffer removals, parallel to a data sub-array activation. In the common case, we remove requests in batches of $M$ that map to the same sub-array row, issuing a single counter sub-array activation for $M$ read-modify-write (RMW) operations. Additionally, a removal can happen in two cases: 
\vspace{0.1in}

\begin{enumerate}
    \item \textbf{Buffer Full:} When the request buffer is full, we trigger eager removal of entries from the RowID with the highest number of entries, to avoid an overflow.
       \item \textbf{RepCount exceeds a threshold:} If any entry's RepCount reaches $K$, we remove it along with other entries for the same row. This is because we reduce the \NBO{} by $K$ to account for the delayed counter access; delaying the counter access beyond $K$ increments would result in loss of timely Rowhammer mitigation and insecurity.
\end{enumerate}

\smallskip
\noindent \textbf{Sizing:} 
We set the request buffer size to 64 per-bank to ensure adequate potential for coalescing counter row activations.
Since $M$ read-modify-writes of 1B counters from the counter sub-array need to occur in the shadow of an 8B data access from the data sub-array for each chip, we set $M=4$. To minimize impact on \NBO{} and rate of \ALERT{}s, we set $K=4$ as this is sufficient for capturing the counter hits in the request buffer.

\subsection{Architecting the Request Buffer} \label{subsec:architect_design}
The design of the request buffer is critical to ensure we can coalesce the most counter row activations. We consider a few different options that balance practicality and the maximum coalescing potential in \Defense{}.

\begin{figure}[htb]
 \centering
\includegraphics[width=0.95\textwidth/2,height=3.8cm,keepaspectratio]  {"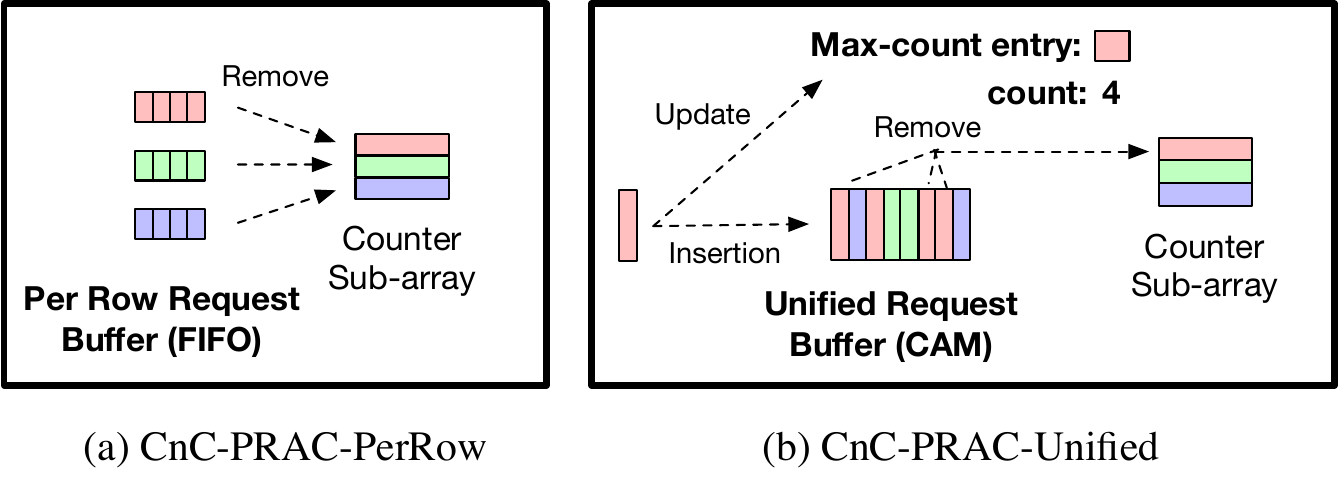"}
  \caption{Designs for the Counter Request Buffer in \Defense{}}
  \label{fig:cnc-design}
\end{figure}

\smallskip
\noindent \textit{Per-Row Request Buffer.} As shown in \cref{fig:cnc-design} (a), one approach is to assign a dedicated buffer to each row in the counter sub-array, with a size equal to the maximum coalescing limit ($M$). \CRS{We call this approach, {\bf \Defense{}-PerRow.}} This design guarantees that $M$ requests can be coalesced, and is highly effective in minimizing activations. However, its storage requirements are $M$ times the number of counter sub-array rows.

\smallskip
\noindent \textit{Unified Request Buffer.} Alternatively, a unified structure with 64 entries for the entire counter sub-array is more storage efficient and desirable. However, a naive unified request buffer with a FCFS-like policy, that looks at the first available RowID in the queue and removes all other entries corresponding to the counter row is inefficient. This is because there may be other rows with a higher number of entries buffered later in the buffer. 
Ideally, we seek to remove the entries corresponding to the counter RowID with the highest number of buffered entries.
A queue sorted based on the counts of entries per RowID can enable this behavior, but insertions become prohibitively expensive, requiring up to $O(N)$ relocations.

To approximate a sorted unified request buffer, we design \CRS{{\bf \Defense{}-Unified}}, as shown in \cref{fig:cnc-design} (b). This approach maintains a CAM while keeping track of just the RowID with the highest count in the buffer and the associated count as metadata. On insertions, the entry count of the inserted RowID is calculated with a single CAM lookup and compared with the maximum entry count, and the metadata is updated. When the entries belonging to the highest count RowID are removed, the maximum count RowID defaults to that of the first entry in the buffer. While this approach captures only a local maximum, we show that it substantially reduces counter row activations.

\begin{figure*}[!htb]
    \centering
\includegraphics[width=6.7in]
{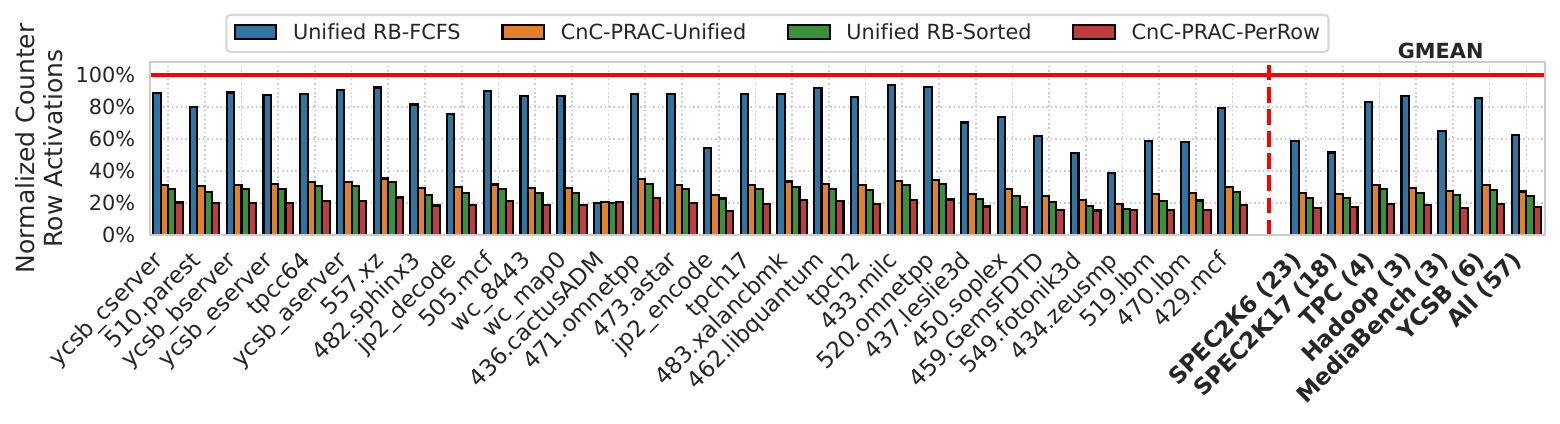}
    \caption{Normalized counter row activations for the different request buffer (RB) designs including Unified RB - FCFS (using the first RowID), \CRS{\Defense{}-Unified}, Unified RB - Sorted (with sorted queues), and \CRS{\Defense{}-PerRow}, normalized to Chronus. }
    \label{fig:normalized_act}
\end{figure*}

\CRS{\subsection{Timings of Coalesced Counter Increments}
\Defense{} coalesces up to $K=4$ counter requests mapping to a single counter row, to enable a single counter-row activation for these requests.
These counters can lie in different columns of a counter-row, requiring multiple column select signals within the counter row.
Fortunately, prior work~\cite{TierdDRAM} has shown that smaller DRAM arrays, such as a 64-row counter-subarray, can have a considerably reduced tRCD compared to a data subarray of 512 rows in DDR3 DRAM.
Thus, multiple ($K$) single byte counter increments are feasible within the shadow of a data row activation; we assume $K=4$. 
A precise evaluation of these timings for DDR5 is left for future work.}

\CRS{
\subsection{Protecting Counter Sub-array from  Bitflips}
To protect the counters themselves from Rowhammer attacks, prior approaches such as introducing guard rows~\cite{van2018guardion,konoth2018zebram} between consecutive rows in the sub-array, or refresh potential victims in parallel to other data row activations\cite{REGA_SP23}, can be adopted in \Defense{} similar to Chronus~\cite{Chronus}. 
Since the counter sub-arrays are small, consisting of just 64 counter rows, these mitigations incur negligible overhead in terms of storage or energy. 
}
\section{Evaluations}

\subsection{Methodology} \label{subsec:method}
\smallskip
\noindent \textbf{Simulation Framework:} We evaluate \Defense{} using the trace-based DRAM simulator Ramulator2~\cite{kim2015ramulator,ramulator2}.
We use an out-of-order core model in Ramulator2, similar to prior RH works~\cite{yauglikcci2021blockhammer, comet, olgun2023abacus, HIRA, rowpress, UPRAC, dapper}. Our system configuration is shown in Table~\ref{table:system_config}. 
We simulate a system with a 4-core, 8MB shared LLC equipped with 32GB DDR5 memory (one channel, two ranks).

\begin{table}[h!]
\vspace{0.1in}
\begin{center}
\begin{small}
\caption{System Configuration}{
\resizebox{0.95\columnwidth}{!}{
\begin{tabular}{|c|c|}
\hline
  Out-Of-Order Cores           &  4 Core, 2GHz, 4 wide, 352 entry ROB          \\
  Last Level Cache (Shared)    & 8MB, 8-Way, 64B lines \\ \hline
  Memory Size, Type                  & 32 GB,  DDR5, x8 chips \\
  Bus Speed             & 1600MHz (3200MHz DDR) \\
  DRAM Organization      & 32 Banks x 2 Ranks x 1 Channel \\
  Rows Per Bank, Size, Size/Chip                 & 64K, 8KB, 8Kbits \\ \hline

\end{tabular}}
\label{table:system_config}
}
\end{small}
\end{center}
\end{table}

\begin{figure*}[!htb]
    \centering
\includegraphics[width=6.7in]{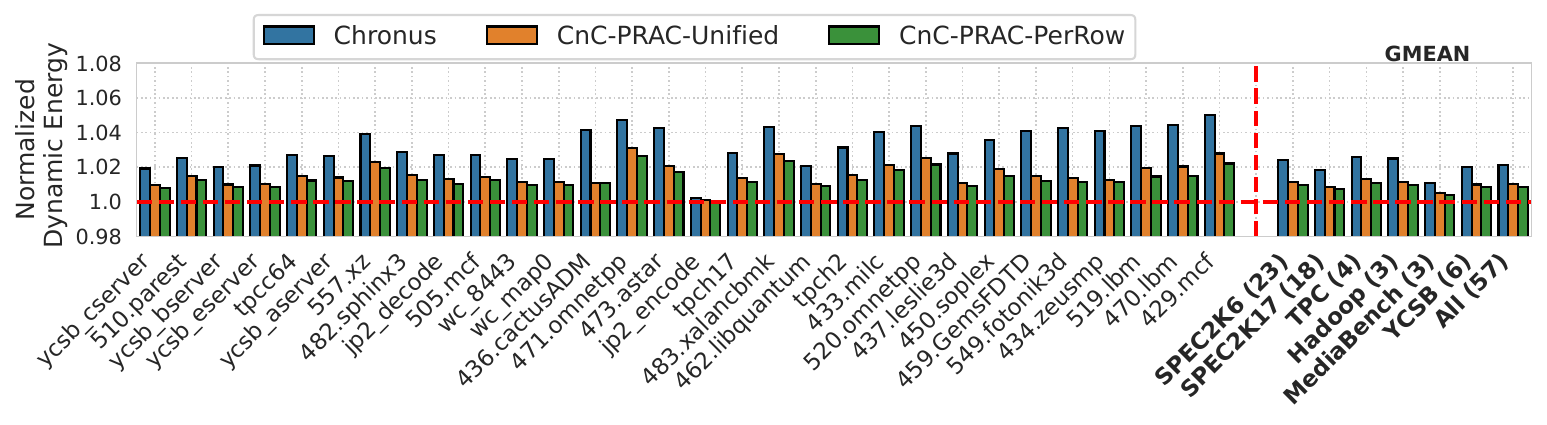}
    \caption{Dynamic energy overhead of Chronus and \Defense{}, normalized to non-secure baseline. 
    Compared to Chronus, which has close to 2.1\% overhead, \CRS{\Defense{}-Unified and \Defense{}-PerRow has only a 1\% and 0.84\% overhead, respectively,} due to fewer counter row activations.}
    \label{fig:energy_overhead}
\end{figure*}

\smallskip
\noindent \textbf{Evaluated Designs}:
We compare \Defense{} against an insecure DRAM baseline and Chronus~\cite{Chronus} (specifically, Chronus-PB).
We additionally compare 4 different request buffer designs: \CRS{\Defense{}-PerRow (with 256 entries, 4 per row), a unified buffer that removes the first  RowID (64 entries), a fully sorted unified buffer (64 entries), and \Defense{}-Unified (64 entries).}  
We use a default Back-Off Threshold (\NBO) of 32 (28 for \Defense{}) and 1 \RFM{} per Alert (PRAC-1).
We also employ a priority-based mitigation service queue, similar to QPRAC, for both Chronus and \Defense{}, allowing one proactive mitigation every 2$\times$\TREFI{}.


\smallskip
\noindent \textbf{Workloads}: We use 57 workloads from SPEC2006~\cite{SPEC2006}, SPEC2017~\cite{SPEC2017}, TPC~\cite{TPC}, Hadoop~\cite{hadoop}, MediaBench~\cite{MediaBench}, and YCSB~\cite{ycsb} provided by Ramulator2~\cite{ramulator_opensource}. We run 4 copies of the workloads (1 per core) until 400 million instructions.

\subsection{Reduction in Counter Row Activations} \label{subsec:eval_counter_access}
\cref{fig:normalized_act} shows the counter subarray activations for the different request buffer designs normalized to Chronus. The naive unified request buffer design, \CRS{Unified RB - FCFS} has almost 62.2\% of Chronus's counter row activations. 
In comparison, \CRS{\Defense{}-Unified} with the approximate maximum entry count per row incurs 27.1\% of the counter row activations as Chronus, nearly identical to an idealized \CRS{Unified RB - Sorted}, with a fully sorted unified request buffer (24.4\%).
This is because \CRS{\Defense{}-Unified}'s removals from the buffer approximate the maximums for the buffered entry counts per row.
Moreover, \CRS{\Defense{}-Unified}'s reductions in counter row activations are comparable to the \CRS{\Defense{}-PerRow} (17.6\%), which always ensures that a single counter row activation is performed for four counter requests.



\subsection{Energy and Performance}

\smallskip
\noindent \textbf{Dynamic Energy:} \Defense{}’s optimization focuses on the dynamic energy overheads, as it aims to reduce the counter sub-array activations. We assume each counter subarray activation, precharge, and one byte Read-Modify-Write (RMW) consumes 19\% extra energy compared to a data row activation, as per Chronus\cite{Chronus}. 
For additional 1-byte RMWs, we estimate it to take $1/8^\text{th}$ of an 8-byte DRAM R/W energy in Ramulator2, after applying a similar ratio as Chronus. 
\cref{fig:energy_overhead} shows the dynamic energy overhead of Chronus and \Defense{} normalized to an insecure baseline. On average, Chronus incurs a 2.1\% dynamic energy overhead, \CRS{while \Defense{}-Unified and \Defense{}-PerRow further reduce this to just 1\% and 0.84\%, respectively}.
These benefits predominantly come from a reduction in counter row activation and precharge operations; we maintain the number of counter read-modify-writes and their energy to be the same.
\CRS{The \Defense{} counter request buffer accesses require less than 1.1 pJ, as reported by Eva-CAM and CACTI~\cite{EvaCAM,CACTI}, which is negligible compared to counter row activation energy.}

\smallskip
\noindent \textbf{Static Energy:}
\Defense's counter sub-array incurs the same background DRAM energy as Chronus~\footnote{Chronus~\cite{Chronus} assumes that the counter sub-array with 64 rows consumes 19\% of the background energy as the data subarray with 128K rows. We leave a more accurate estimation of the background energy for future work.}. 
\CRS{
In \Defense{}-Unifed, the request buffer requires a 64-entry CAM, similar in size as a TRR tracker in DDR4~\cite{hassan2021uncovering}, and smaller than prior in-DRAM trackers~\cite{ProTRR, park2020graphene}.  
Using Eva-CAM\cite{EvaCAM} with a 45nm process\cite{EVACAM_CELL}, we estimate it to incur 4.3 mW of static power per DRAM chip.
In comparison, the \Defense{}-PerRow incurs a static power of 0.25 mW per DRAM chip, as reported by CACTI~\cite{CACTI}, around 20$\times$ less than \Defense{}-Unified.
While the per-row design is beneficial in terms of static power, it is less efficient in storage.
The unified design with 64 entries requires 192 bytes per bank (3 bytes per entry), whereas the per-row design with 256 entries (4 entries per row) requires 384 bytes per bank with 12 bits per entry (10-bit byte-position, 2-bit RepCount).
We leave the exploration of designs that incur both low static energy and storage for future work as this requires more accurate SPICE-modeling of counter sub-arrays and request buffer designs.}

\smallskip
\noindent \textbf{Performance.} On average, we observe that both Chronus and \Defense{} have negligible performance overheads (below 0.5\%), when combined with proactive mitigations on \TREFI{}s, as enabled by QPRAC's priority mitigation service queues.

\subsection{Sensitivity to Request Buffer Size}
\begin{figure}[htb]
\centering
\includegraphics[width=3.4in,height=\paperheight,keepaspectratio]{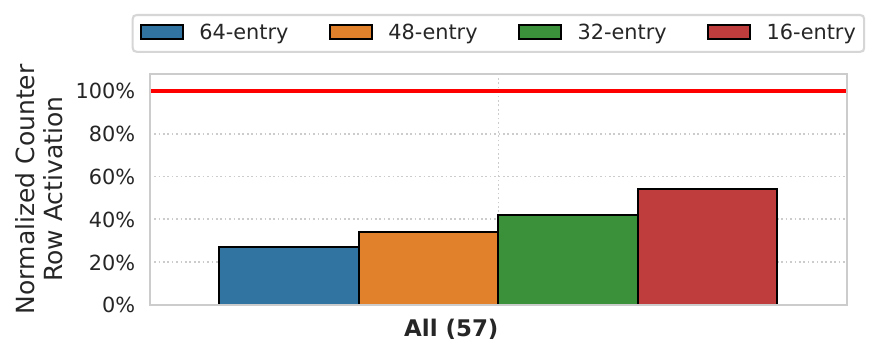}
\caption{Sensitivity of \CRS{\Defense{}-Unified's} counter row activations as the request buffer size varies. Reducing to 1/4 of the original size results in an increase of normalized activation from 27.1\% to 74.7\%.}
\label{fig:buffer_size_sensitivity}
\end{figure}

A smaller request buffer provides less reduction in counter row activation counts. 
As shown in \cref{fig:buffer_size_sensitivity}, varying the request buffer size \CRS{of \Defense{}-Unified} from 64 to 16 causes the counter row activation counts to increase from 27\% to 74\% of the activations in Chronus.
This corresponds to dynamic energy overhead increasing from 1\% to 1.8\% compared to a non-secure baseline.


\section{Why Not Cache Row Activation Counts?}
\Defense{} advocates for coalescing counter row requests to the same row to minimize counter row activations. An orthogonal strategy is to introduce an in-DRAM counter cache to reduce counter accesses from the DRAM arrays. 
We characterize the row activation patterns across all DRAM rows for workloads, and identify that for capturing even 50\% of the counter row activations, the counter footprint would cross almost 600 to 2,000 to 7,500 in small, medium or large memory footprint workloads, as shown in \cref{fig:row_dist} in the Appendix. 

Consequently, for a counter cache to be effective it would require 1000s of entries per bank. 
Such large SRAM-based caches can be prohibitive to incorporate within DRAM, which can often only accommodate 50-100 entries per bank.  
We evaluate the LRU-based caches with approximately 100 entries per DRAM Chip and observe that such caches do not provide significant reuse and do not meaningfully reduce the row activation counts. 
Future works can consider better filtering or caching techniques for activation counters in DRAM chips.

\section{Related Work}
\noindent{\bf Prior PRAC Designs:} PRAC was inspired by Panopticon, which proposed per row activation counters. MOAT~\cite{MOAT} and QPRAC~\cite{QPRAC} showed vulnerabilities in Panopticon due to tardy mitigations and FIFO queues and proposed efficient defenses. Chronus~\cite{Chronus} further tackled the performance overheads of PRAC's DRAM timings by using a separate counter sub-array for counter updates. Building on Chronus, we further reduce energy overheads by coalescing counter row activations.

\noindent{\bf Hybrid Mitigation:} Hydra~\cite{qureshi2022hydra} uses per-row counters in DRAM with SRAM filtering and caching in the memory controller. Still, its Group Counting Tables (1000 entries per bank) and cache (8000 entries per channel) incur impractical storage overhead. Our design instead coalesces counter accesses, achieving energy efficiency without prohibitive storage.

\noindent{\bf Efficient Aggressor Counting:} SRAM-based trackers like CAT~\cite{CBT}, TWiCE~\cite{lee2019twice}, and Mithril~\cite{kim2022mithril} are storage-intensive and impractical at sub-100 \TRH{}. Probabilistic trackers, such as DSAC~\cite{DSAC} and PAT~\cite{HynixRH}, are insecure or unreliable, while PrIDE~\cite{jaleel2024pride} suffers significant bandwidth loss at low \TRH{}. Unlike these approaches, our design ensures deterministic security with minimal storage and negligible overheads.

\begin{figure*}[ht]
\vspace{0.2in}
\centering
\begin{subfigure}[t]{0.32\linewidth}
\centering
\includegraphics[width=2.2in,height=\paperheight,keepaspectratio]{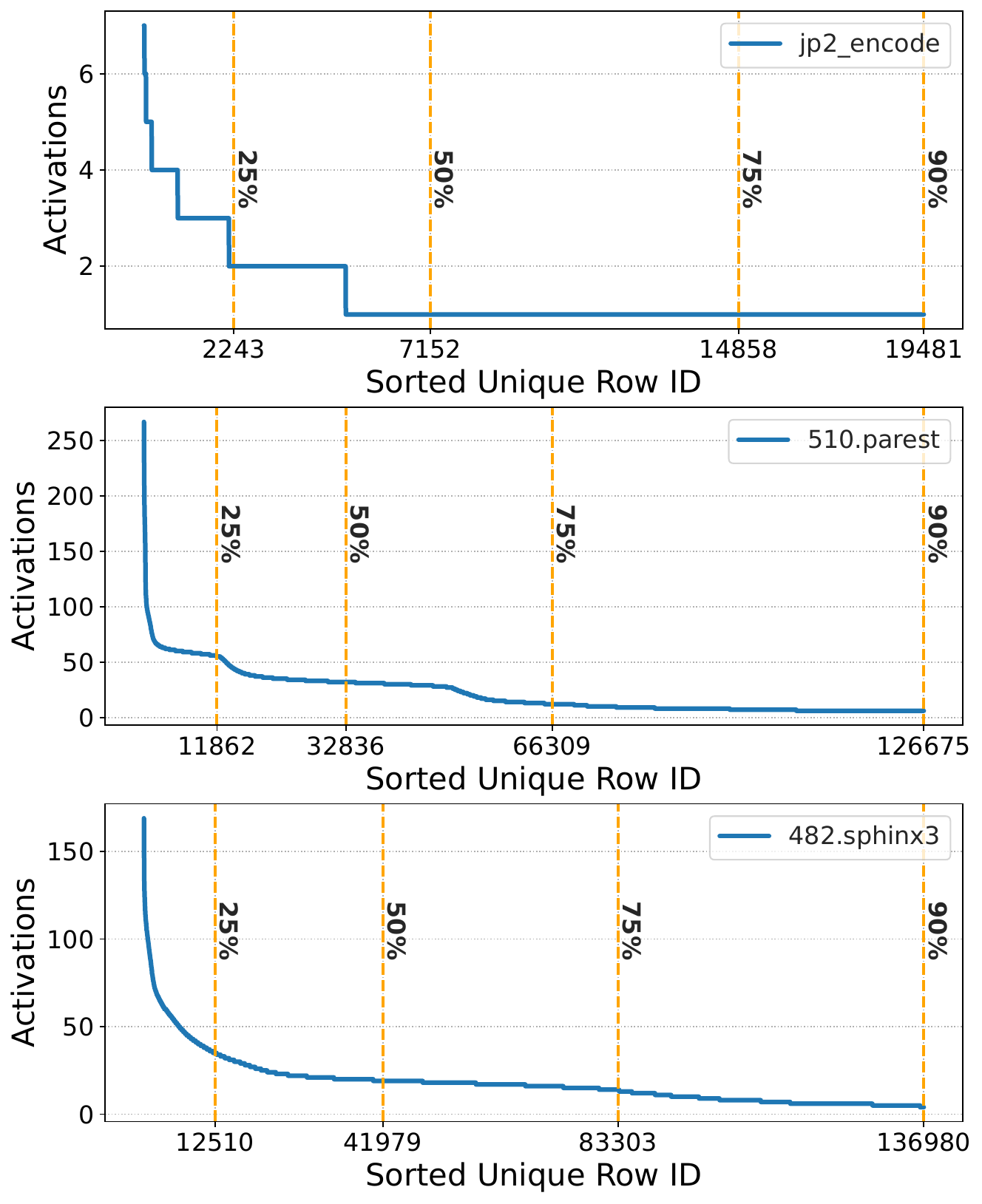}
\caption{Small Footprint}
\label{fig:row_dist_small}
\end{subfigure}
\begin{subfigure}[t]{.32\linewidth}
\centering
\includegraphics[width=2.2in,height=\paperheight,keepaspectratio]{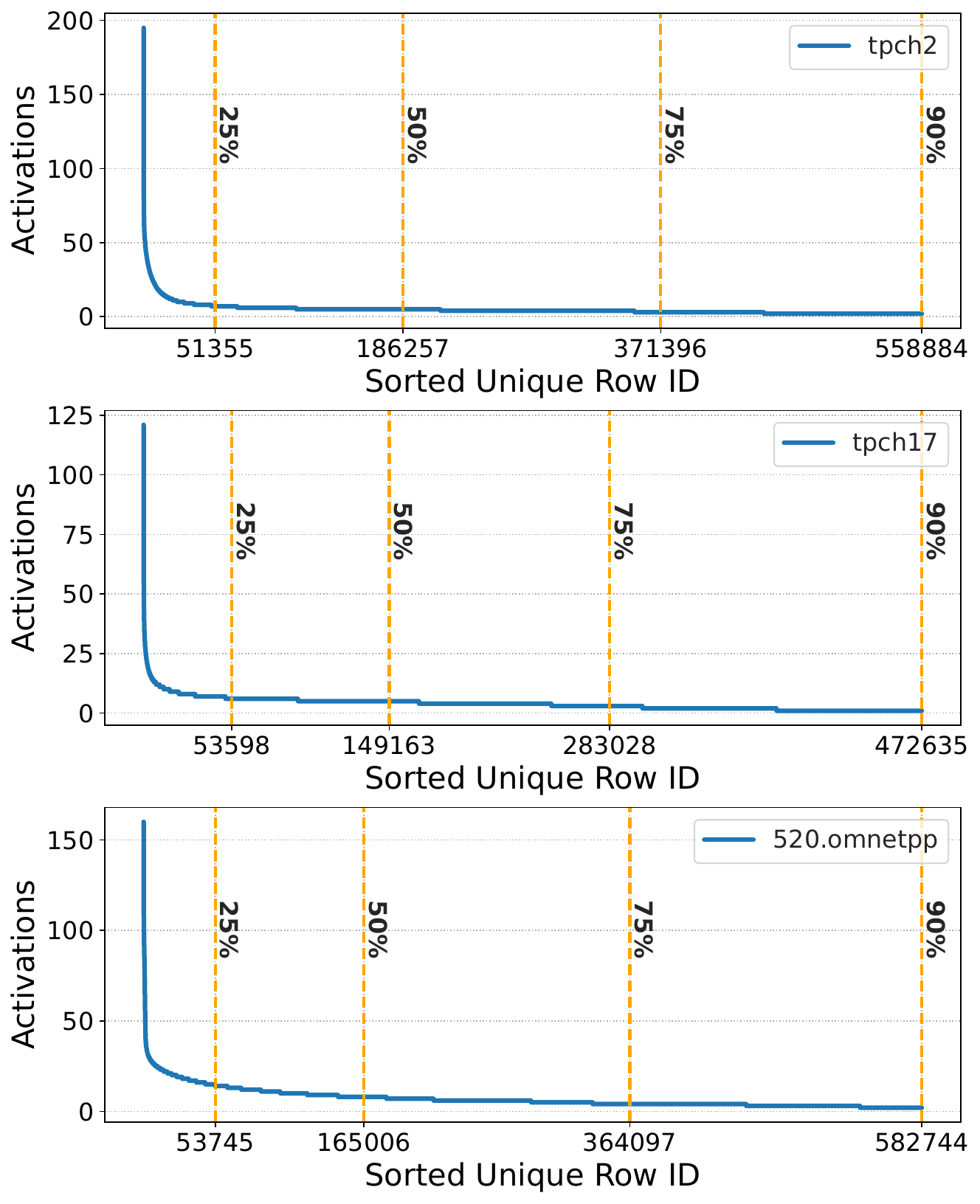}
\caption{Medium Footprint}
\label{fig:row_dist_medium}
\end{subfigure}
\begin{subfigure}[t]{.32\linewidth}
\centering
\includegraphics[width=2.2in,height=\paperheight,keepaspectratio]{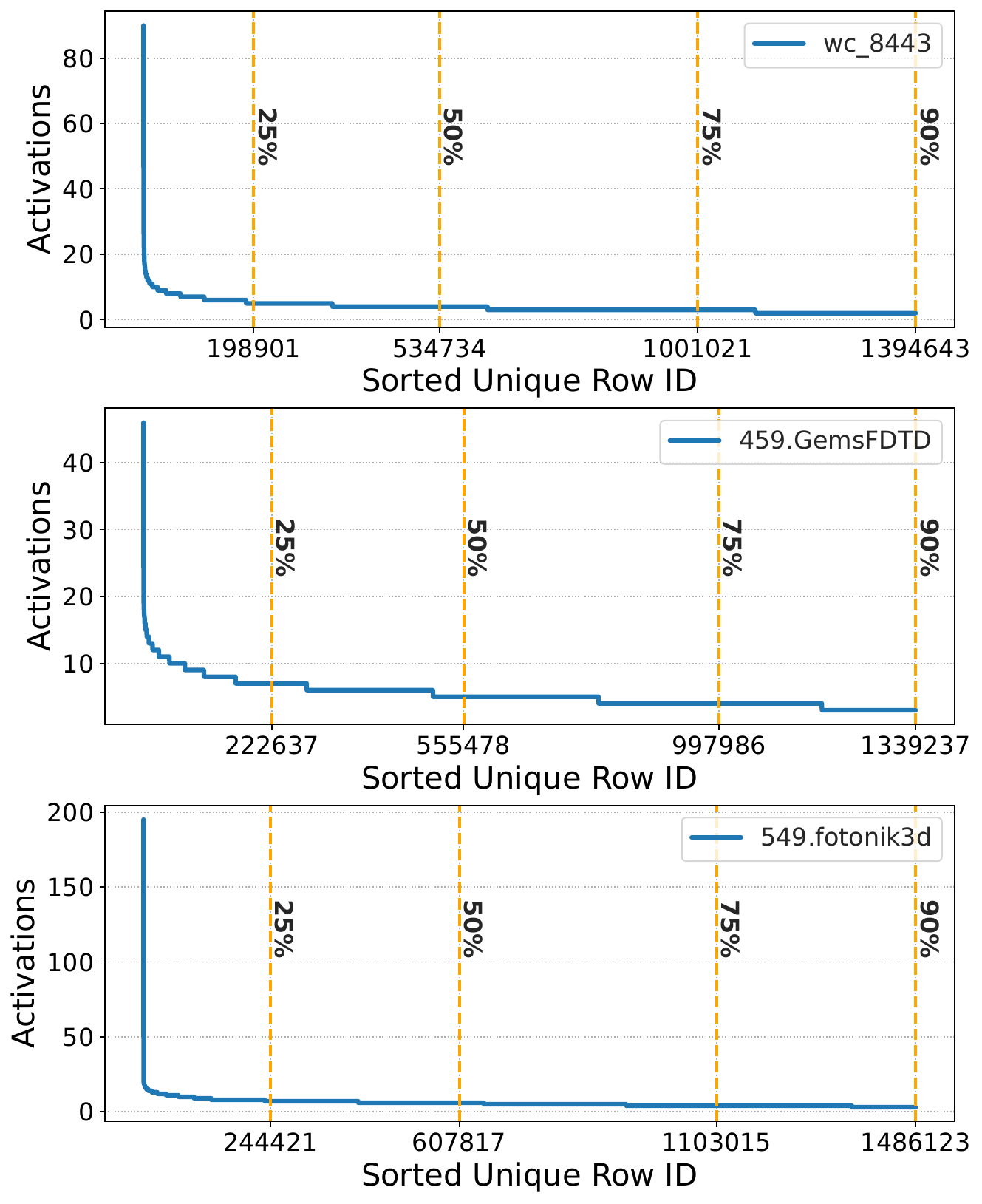}
\caption{Large Footprint}
\label{fig:row_dist_large}
\vspace{0.2in}
\end{subfigure}

\caption{Activations to unique DRAM rows across 64 banks, sorted by activation count. We classify workloads into (a) small, (b) medium, and (c) large memory footprints and illustrate three workloads in each category. The number of rows per DRAM bank accounting for 50\% of total activations increases with footprint size, ranging from 600 per bank in the small footprint workload to 2,000 and 7,500 per bank in the medium and large ones, respectively.}
\label{fig:row_dist}

\end{figure*}

\section{Conclusion}
This paper introduces \Defense{}, a novel PRAC implementation that enables an energy-efficient Rowhammer mitigation. By reordering and coalescing accesses to counters within the same row, and decoupling counter operations from the critical path of data accesses, our design reduces counter-related row activations by \CRS{75\%-83\%}. This enables negligible slowdown and a minimal \CRS{0.84\%-1\%} energy overhead compared to baseline DDR5 DRAM, surpassing state-of-the-art solutions.

\appendix
\begin{appendices}

\section*{Appendix A: DRAM Row ACTs Distribution}
\cref{fig:row_dist} shows the distribution of activations per DRAM row across the entire memory, with the rows sorted in descending order of activations. We classify the workloads based on their memory footprint sizes into small, medium and large. The dotted yellow lines indicate the 25, 50, 75, and 90 percentile rows in terms of total activations. 

In a small footprint workload like \textit{jp2\_encode}, 7,152 rows across the memory (111 rows per bank) account for 50\% of total activations. This skew suggests that a small subset of rows receives disproportionately high access frequency, offering an opportunity for caching to effectively capture a reasonable fraction of the counter accesses. 
However, as the memory footprint increases, the number of heavily accessed rows also grows.
For example, in a medium footprint workload like tpch2, 186,257 rows across memory (2910 rows per bank) account for 50\% of activations. 
Similarly, in a large footprint workload like wc\_8443, 534,734 rows across memory (8355 rows per bank) account for 50\% of activations. 
This makes it difficult for resource-constrained caches to maintain effective hit rates and provide benefits in larger workloads.

\section*{Appendix-B: Evaluation of Caches}
To evaluate the impact of caches, we extend  \Defense{} and add an in-DRAM byte-level cache that caches 1B counters alongwith identifier metadata (tag, and dirty/valid bits).
Whenever a read-modify-write is serviced from the request queue, we install a clean copy of the counter in the cache. 
If there are any writebacks due to dirty evictions on a cache insertion, they are inserted into the counter request buffer. We re-purpose one bit of the RepCount to differentiate a write-back from a RMW request in the request buffer. 
The goal of the cache is to ensure that hits for counter accesses can completely eliminate the counter row activation and read/write access for the counter.

\begin{figure}[htb]
    \centering
\includegraphics[width=2in]
{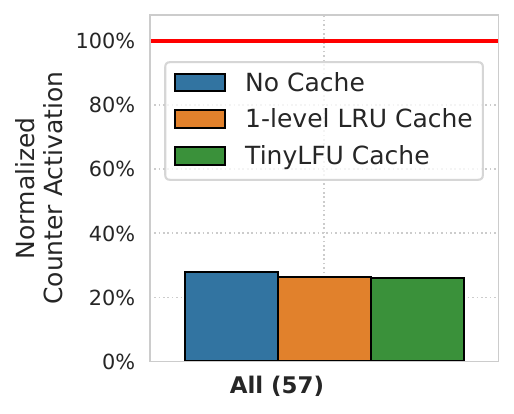}
    \caption{Counter row activations with \Defense{} without and with caching, normalized to Chronus. Across all caching-based designs, the reduction in counter row activations is only 1.2\%, primarily due to the small cache size relative to the large number of frequently activated rows(see \cref{fig:row_dist}).}
    \label{fig:cache_act_comparison}
\end{figure}

We prototyped two cache designs: (1) a 4-way set-associative cache with LRU replacement and (2) a 2-level admission-based cache inspired by TinyLFU \cite{einziger2017tinylfu}. The admission-based cache is designed to avoid polluting the cache with low-value entries, particularly effective under long-tailed access patterns. Both caches are limited to 64 entries due to storage constraints. 
The caches, however, bring minimal improvement in reducing counter row activations as seen in \cref{fig:cache_act_comparison}.
This is mainly because of the limited size of the cache is unable to capture many hits, showing a hit rate of less than 3.5\%.
Future works can explore better caching techniques within the constraints of the limited storage available in-DRAM. 

\end{appendices}


\balance
\bibliographystyle{IEEEtran}
\bibliography{refs}

\begin{thebibliography}{10}
\providecommand{\url}[1]{#1}
\csname url@samestyle\endcsname
\providecommand{\newblock}{\relax}
\providecommand{\bibinfo}[2]{#2}
\providecommand{\BIBentrySTDinterwordspacing}{\spaceskip=0pt\relax}
\providecommand{\BIBentryALTinterwordstretchfactor}{4}
\providecommand{\BIBentryALTinterwordspacing}{\spaceskip=\fontdimen2\font plus
\BIBentryALTinterwordstretchfactor\fontdimen3\font minus \fontdimen4\font\relax}
\providecommand{\BIBforeignlanguage}[2]{{%
\expandafter\ifx\csname l@#1\endcsname\relax
\typeout{** WARNING: IEEEtran.bst: No hyphenation pattern has been}%
\typeout{** loaded for the language `#1'. Using the pattern for}%
\typeout{** the default language instead.}%
\else
\language=\csname l@#1\endcsname
\fi
#2}}
\providecommand{\BIBdecl}{\relax}
\BIBdecl

\bibitem{kim2014flipping}
Y.~Kim, R.~Daly, J.~Kim, C.~Fallin, J.~H. Lee, D.~Lee, C.~Wilkerson, K.~Lai, and O.~Mutlu, ``Flipping bits in memory without accessing them: An experimental study of dram disturbance errors,'' \emph{ISCA}, 2014.

\bibitem{seaborn2015exploiting}
M.~Seaborn and T.~Dullien, ``{Exploiting the DRAM rowhammer bug to gain kernel privileges},'' \emph{Black Hat}, vol.~15, p.~71, 2015.

\bibitem{frigo2020trrespass}
P.~Frigo, E.~Vannacc, H.~Hassan, V.~Van Der~Veen, O.~Mutlu, C.~Giuffrida, H.~Bos, and K.~Razavi, ``{TRRespass: Exploiting the many sides of target row refresh},'' in \emph{IEEE Symposium on Security and Privacy}, 2020.

\bibitem{gruss2018another}
D.~Gruss, M.~Lipp, M.~Schwarz, D.~Genkin, J.~Juffinger, S.~O'Connell, W.~Schoechl, and Y.~Yarom, ``Another flip in the wall of rowhammer defenses,'' in \emph{IEEE Symposium on Security and Privacy}, 2018.

\bibitem{aweke2016anvil}
Z.~B. Aweke, S.~F. Yitbarek, R.~Qiao, R.~Das, M.~Hicks, Y.~Oren, and T.~Austin, ``Anvil: Software-based protection against next-generation rowhammer attacks,'' in \emph{ASPLOS}, 2016.

\bibitem{cojocar2019eccploit}
L.~Cojocar, K.~Razavi, C.~Giuffrida, and H.~Bos, ``Exploiting correcting codes: On the effectiveness of ecc memory against rowhammer attacks,'' in \emph{IEEE Symposium on Security and Privacy (SP)}, 2019.

\bibitem{gruss2016rhjs}
D.~Gruss, C.~Maurice, and S.~Mangard, ``Rowhammer.js: A remote software-induced fault attack in javascript,'' in \emph{Detection of Intrusions and Malware, and Vulnerability Assessment}, 2016.

\bibitem{vanderveen2016drammer}
V.~van~der Veen, Y.~Fratantonio, M.~Lindorfer, D.~Gruss, C.~Maurice, G.~Vigna, H.~Bos, K.~Razavi, and C.~Giuffrida, ``Drammer: Deterministic rowhammer attacks on mobile platforms,'' in \emph{ACM-CCS}, 2016.

\bibitem{kim2020revisitingRH}
J.~S. Kim, M.~Patel, A.~G. Ya{\u{g}}l{\i}k{\c{c}}{\i}, H.~Hassan, R.~Azizi, L.~Orosa, and O.~Mutlu, ``Revisiting rowhammer: An experimental analysis of modern dram devices and mitigation techniques,'' in \emph{ISCA}, 2020.

\bibitem{jedec_ddr5_prac}
\mbox JEDEC. {JESD79-5C}. \url{https://www.jedec.org/document_search?search_api_views_fulltext=jesd79-5c}.

\bibitem{MOAT}
M.~Qureshi and S.~Qazi, ``{MOAT: Securely Mitigating Rowhammer with Per-Row Activation Counters},'' in \emph{Proceedings of the 30th ACM International Conference on Architectural Support for Programming Languages and Operating Systems (ASPLOS)}, 2025.

\bibitem{QPRAC}
J.~Woo, S.~C. Lin, P.~J. Nair, A.~Jaleel, and G.~Saileshwar, ``{QPRAC: Towards secure and practical prac-based rowhammer mitigation using priority queues},'' in \emph{2025 IEEE International Symposium on High Performance Computer Architecture (HPCA)}.\hskip 1em plus 0.5em minus 0.4em\relax IEEE, 2025, pp. 1021--1037.

\bibitem{Chronus}
O.~Canpolat, A.~G. Ya{\u{g}}l{\i}k{\c{c}}{\i}, G.~F. Oliveira, A.~Olgun, N.~Bostanc{\i}, I.~E. Yuksel, H.~Luo, O.~Ergin, and O.~Mutlu, ``Chronus: Understanding and securing the cutting-edge industry solutions to dram read disturbance,'' in \emph{2025 IEEE International Symposium on High Performance Computer Architecture (HPCA)}.\hskip 1em plus 0.5em minus 0.4em\relax IEEE, 2025, pp. 887--905.

\bibitem{SALP}
Y.~Kim, V.~Seshadri, D.~Lee, J.~Liu, and O.~Mutlu, ``A case for exploiting subarray-level parallelism (salp) in dram,'' in \emph{2012 39th Annual International Symposium on Computer Architecture (ISCA)}, 2012, pp. 368--379.

\bibitem{bennett2021panopticon}
T.~Bennett, S.~Saroiu, A.~Wolman, and L.~Cojocar, ``Panopticon: A complete in-dram rowhammer mitigation,'' in \emph{Workshop on DRAM Security (DRAMSec)}, 2021.

\bibitem{TierdDRAM}
D.~Lee, Y.~Kim, V.~Seshadri, J.~Liu, L.~Subramanian, and O.~Mutlu, ``Tiered-latency dram: A low latency and low cost dram architecture,'' in \emph{2013 IEEE 19th International Symposium on High Performance Computer Architecture (HPCA)}, 2013, pp. 615--626.

\bibitem{van2018guardion}
V.~Van~der Veen, M.~Lindorfer, Y.~Fratantonio, H.~P. Pillai, G.~Vigna, C.~Kruegel, H.~Bos, and K.~Razavi, ``Guardion: Practical mitigation of dma-based rowhammer attacks on arm,'' in \emph{International Conference on Detection of Intrusions and Malware, and Vulnerability Assessment}.\hskip 1em plus 0.5em minus 0.4em\relax Springer, 2018, pp. 92--113.

\bibitem{konoth2018zebram}
R.~K. Konoth, M.~Oliverio, A.~Tatar, D.~Andriesse, H.~Bos, C.~Giuffrida, and K.~Razavi, ``{ZebRAM: comprehensive and compatible software protection against rowhammer attacks},'' in \emph{13th USENIX - (OSDI 18)}, 2018, pp. 697--710.

\bibitem{REGA_SP23}
M.~Marazzi, F.~Solt, P.~Jattke, K.~Takashi, and K.~Razavi, ``{REGA: Scalable Rowhammer Mitigation with Refresh-Generating Activations},'' in \emph{{IEEE Symposium on Security and Privacy (SP)}}.\hskip 1em plus 0.5em minus 0.4em\relax IEEE, 2023.

\bibitem{kim2015ramulator}
Y.~Kim, W.~Yang, and O.~Mutlu, ``Ramulator: A fast and extensible dram simulator,'' \emph{IEEE Computer architecture letters}, vol.~15, no.~1, pp. 45--49, 2015.

\bibitem{ramulator2}
H.~Luo, Y.~C. Tuğrul, F.~N. Bostancı, A.~Olgun, A.~G. Yağlıkçı, and O.~Mutlu, ``Ramulator 2.0: A modern, modular, and extensible dram simulator,'' \emph{IEEE Computer Architecture Letters}, vol.~23, no.~1, pp. 112--116, 2024.

\bibitem{yauglikcci2021blockhammer}
A.~G. Ya{\u{g}}lik{\c{c}}i \emph{et~al.}, ``Blockhammer: Preventing rowhammer at low cost by blacklisting rapidly-accessed dram rows,'' in \emph{HPCA}, 2021.

\bibitem{comet}
F.~N. Bostanci, I.~E. Yüksel, A.~Olgun, K.~Kanellopoulos, Y.~C. Tuğrul, A.~G. Yağliçi, M.~Sadrosadati, and O.~Mutlu, ``Comet: Count-min-sketch-based row tracking to mitigate rowhammer at low cost,'' in \emph{2024 IEEE International Symposium on High-Performance Computer Architecture (HPCA)}, 2024, pp. 593--612.

\bibitem{olgun2023abacus}
\BIBentryALTinterwordspacing
A.~Olgun, Y.~C. Tugrul, N.~Bostanci, I.~E. Yuksel, H.~Luo, S.~Rhyner, A.~G. Yaglikci, G.~F. Oliveira, and O.~Mutlu, ``{ABACuS}: {All-Bank} activation counters for scalable and low overhead {RowHammer} mitigation,'' in \emph{33rd USENIX Security Symposium (USENIX Security 24)}.\hskip 1em plus 0.5em minus 0.4em\relax Philadelphia, PA: USENIX Association, Aug. 2024, pp. 1579--1596. [Online]. Available: \url{https://www.usenix.org/conference/usenixsecurity24/presentation/olgun}
\BIBentrySTDinterwordspacing

\bibitem{HIRA}
A.~G. Yağlikçi, A.~Olgun, M.~Patel, H.~Luo, H.~Hassan, L.~Orosa, O.~Ergin, and O.~Mutlu, ``Hira: Hidden row activation for reducing refresh latency of off-the-shelf dram chips,'' in \emph{MICRO}, 2022.

\bibitem{rowpress}
H.~Luo, A.~Olgun, A.~G. Ya\u{g}l\i{}k\c{c}\i{}, Y.~C. Tu\u{g}rul, S.~Rhyner, M.~B. Cavlak, J.~Lindegger, M.~Sadrosadati, and O.~Mutlu, ``Rowpress: Amplifying read disturbance in modern dram chips,'' in \emph{ISCA-50}, 2023.

\bibitem{UPRAC}
O.~Canpolat, A.~G. Ya{\u{g}}l{\i}k{\c{c}}{\i}, G.~F. Oliveira, A.~Olgun, O.~Ergin, and O.~Mutlu, ``Understanding the security benefits and overheads of emerging industry solutions to dram read disturbance,'' in \emph{Workshop on DRAM Security (DRAMSec)}, 2024.

\bibitem{dapper}
J.~Woo and P.~J. Nair, ``Dapper: A performance-attack-resilient tracker for rowhammer defense,'' in \emph{2025 IEEE International Symposium on High-Performance Computer Architecture (HPCA)}, 2025.

\bibitem{SPEC2006}
\BIBentryALTinterwordspacing
S.~P.~E. Corporation, ``Spec cpu2006 benchmark suite,'' 2006. [Online]. Available: \url{http://www.spec.org/cpu2006/}
\BIBentrySTDinterwordspacing

\bibitem{SPEC2017}
\BIBentryALTinterwordspacing
``{SPEC CPU2017 Benchmark Suite},'' Standard Performance Evaluation Corporation. [Online]. Available: \url{http://www.spec.org/cpu2017/}
\BIBentrySTDinterwordspacing

\bibitem{TPC}
\BIBentryALTinterwordspacing
{Transaction Processing Performance Council}, ``{TPC Benchmarks}.'' [Online]. Available: \url{http://tpc.org/}
\BIBentrySTDinterwordspacing

\bibitem{hadoop}
\BIBentryALTinterwordspacing
A.~Foundation, ``Apache hadoop.'' [Online]. Available: \url{http://hadoop.apache.org/}
\BIBentrySTDinterwordspacing

\bibitem{MediaBench}
J.~E. Fritts, F.~W. Steiling, J.~A. Tucek, and W.~Wolf, ``Mediabench ii video: Expediting the next generation of video systems research,'' \emph{Microprocessors and Microsystems}, vol.~33, no.~4, pp. 301--318, 2009.

\bibitem{ycsb}
B.~F. Cooper, A.~Silberstein, E.~Tam, R.~Ramakrishnan, and R.~Sears, ``Benchmarking cloud serving systems with ycsb,'' in \emph{Proceedings of the 1st ACM symposium on Cloud computing}, 2010, pp. 143--154.

\bibitem{ramulator_opensource}
\BIBentryALTinterwordspacing
{SAFARI Research Group}, ``{ABACuS — GitHub Repository},'' 2023. [Online]. Available: \url{https://github.com/CMU-SAFARI/ABACuS}
\BIBentrySTDinterwordspacing

\bibitem{EvaCAM}
L.~Liu, M.~M. Sharifi, R.~Rajaei, A.~Kazemi, K.~Ni, X.~Yin, M.~Niemier, and X.~S. Hu, ``Eva-cam: A circuit/architecture-level evaluation tool for general content addressable memories,'' in \emph{2022 Design, Automation \& Test in Europe Conference \& Exhibition (DATE)}, 2022, pp. 1173--1176.

\bibitem{CACTI}
R.~Balasubramonian, A.~B. Kahng, N.~Muralimanohar, A.~Shafiee, and V.~Srinivas, ``Cacti 7: New tools for interconnect exploration in innovative off-chip memories,'' \emph{ACM Transactions on Architecture and Code Optimization (TACO)}, vol.~14, no.~2, pp. 1--25, 2017.

\bibitem{hassan2021uncovering}
H.~Hassan, Y.~C. Tugrul, J.~S. Kim, V.~Van~der Veen, K.~Razavi, and O.~Mutlu, ``Uncovering in-dram rowhammer protection mechanisms: A new methodology, custom rowhammer patterns, and implications,'' in \emph{MICRO-54}, 2021, pp. 1198--1213.

\bibitem{ProTRR}
M.~Marazzi, P.~Jattke, F.~Solt, and K.~Razavi, ``{Protrr: Principled yet optimal in-dram target row refresh},'' in \emph{{IEEE Symposium on Security and Privacy (SP)}}.\hskip 1em plus 0.5em minus 0.4em\relax IEEE, 2022, pp. 735--753.

\bibitem{park2020graphene}
Y.~Park, W.~Kwon, E.~Lee, T.~J. Ham, J.~H. Ahn, and J.~W. Lee, ``Graphene: Strong yet lightweight row hammer protection,'' in \emph{MICRO}.\hskip 1em plus 0.5em minus 0.4em\relax IEEE, 2020, pp. 1--13.

\bibitem{EVACAM_CELL}
X.~Yin, K.~Ni, D.~Reis, S.~Datta, M.~Niemier, and X.~S. Hu, ``An ultra-dense 2fefet tcam design based on a multi-domain fefet model,'' \emph{IEEE Transactions on Circuits and Systems II: Express Briefs}, vol.~66, no.~9, pp. 1577--1581, 2019.

\bibitem{qureshi2022hydra}
M.~Qureshi, A.~Rohan, G.~Saileshwar, and P.~J. Nair, ``Hydra: enabling low-overhead mitigation of row-hammer at ultra-low thresholds via hybrid tracking,'' in \emph{ISCA}, 2022.

\bibitem{CBT}
S.~M. Seyedzadeh, A.~K. Jones, and R.~Melhem, ``Mitigating wordline crosstalk using adaptive trees of counters,'' in \emph{ISCA}, 2018.

\bibitem{lee2019twice}
E.~Lee, I.~Kang, S.~Lee, G.~E. Suh, and J.~H. Ahn, ``{TWiCe: preventing row-hammering by exploiting time window counters},'' in \emph{ISCA}, 2019.

\bibitem{kim2022mithril}
M.~J. Kim, J.~Park, Y.~Park, W.~Doh, N.~Kim, T.~J. Ham, J.~W. Lee, and J.~H. Ahn, ``Mithril: Cooperative row hammer protection on commodity dram leveraging managed refresh,'' in \emph{HPCA}, 2022.

\bibitem{DSAC}
S.~Hong, D.~Kim, J.~Lee, R.~Oh, C.~Yoo, S.~Hwang, and J.~Lee, ``Dsac: Low-cost rowhammer mitigation using in-dram stochastic and approximate counting algorithm,'' \emph{arXiv:2302.03591}, 2023.

\bibitem{HynixRH}
W.~Kim, C.~Jung, S.~Yoo, D.~Hong, J.~Hwang, J.~Yoon, O.~Jung, J.~Choi, S.~Hyun, M.~Kang, S.~Lee, D.~Kim, S.~Ku, D.~Choi, N.~Joo, S.~Yoon, J.~Noh, B.~Go, C.~Kim, S.~Hwang, M.~Hwang, S.-M. Yi, H.~Kim, S.~Heo, Y.~Jang, K.~Jang, S.~Chu, Y.~Oh, K.~Kim, J.~Kim, S.~Kim, J.~Hwang, S.~Park, J.~Lee, I.~Jeong, J.~Cho, and J.~Kim, ``A 1.1v 16gb ddr5 dram with probabilistic-aggressor tracking, refresh-management functionality, per-row hammer tracking, a multi-step precharge, and core-bias modulation for security and reliability enhancement,'' in \emph{ISSCC}, 2023.

\bibitem{jaleel2024pride}
A.~Jaleel, G.~Saileshwar, S.~Keckler, and M.~Qureshi, ``Pride: Achieving secure rowhammer mitigation with low-cost in-dram trackers,'' in \emph{Annual International Symposium on Computer Architecture}, 2024.

\bibitem{einziger2017tinylfu}
G.~Einziger, R.~Friedman, and B.~Manes, ``Tinylfu: A highly efficient cache admission policy,'' \emph{ACM Transactions on Storage (ToS)}, vol.~13, no.~4, pp. 1--31, 2017.

\end{thebibliography}
\balance
\vspace{12pt}
\end{document}